\documentclass[%
reprint,
superscriptaddress,
 amsmath,amssymb,
 aps,
]{revtex4-1}

\usepackage{graphicx}
\usepackage{dcolumn}
\usepackage{bm}

\begin{document}


\title{Monitoring the build-up of hydrogen polarization for polarized 
Hydrogen--Deuteride (HD) targets with NMR at 17 Tesla }%

\newcommand{\TOKYOADDRESS}{Department of Radiology, The University of Tokyo 
Hospital, Tokyo 113-8655, Japan}
\newcommand{\RCNPADDRESS}{Research Center for Nuclear Physics, Osaka University, 
Ibaraki, Osaka 567-0047, Japan}
\newcommand{\NAGOYAADDRESS}{Nagoya University, Chikusa-ku, Nagoya, Aichi, 
464-8602, Japan}
\newcommand{\SINICAADDRESS}{Institute of Physics, Academia Sinica, Taipei 11529, 
Taiwan}
\newcommand{\QSTADDRESS}{National Institutes for Quantum and Radiological 
Science and Technology, Tokai, Ibaraki 319-1195, Japan}

\author{T.~Ohta}\affiliation{\TOKYOADDRESS}\affiliation{\RCNPADDRESS}
\author{M.~Fujiwara}\affiliation{\RCNPADDRESS}\affiliation{\QSTADDRESS}
\author{T.~Hotta}\affiliation{\RCNPADDRESS}
\author{I.~Ide}\affiliation{\NAGOYAADDRESS}
\author{K.~Ishizaki}\affiliation{\NAGOYAADDRESS}
\author{H.~Kohri}\affiliation{\RCNPADDRESS}\affiliation{\NAGOYAADDRESS}\affiliation{\SINICAADDRESS}
\author{Y.~Yanai}\affiliation{\RCNPADDRESS}
\author{M.~Yosoi}\affiliation{\RCNPADDRESS}

\date{\today}

\begin{abstract}
We report on the frozen-spin polarized hydrogen--deuteride (HD) targets for photoproduction 
experiments at SPring-8/LEPS. 
Pure HD gas with a small amount of ortho-H$_{2}$ ($\sim$0.1\%) was liquefied and solidified 
by liquid helium. 
The temperature of the produced solid HD was reduced to about 30 mK with a dilution refrigerator. 
A magnetic field (17 T) was applied to the HD to grow the polarization with the static method. 
After the aging of the HD at low temperatures in the presence of a high-magnetic field strength 
for 3 months, the polarization froze. 
Almost all ortho-H$_{2}$ molecules were converted to para-H$_{2}$ molecules that exhibited 
weak spin interactions with the HD. 
If the concentration of the ortho-H$_{2}$ was reduced at the beginning of the aging process, 
the aging time can be shortened. 
We have developed a new nuclear magnetic resonance (NMR) system to measure the relaxation times ($T_{1}$) 
of the $^{1}$H and $^{2}$H nuclei with two frequency sweeps at the respective frequencies of 726 and 111 MHz, 
and succeeded in the monitoring of the polarization build-up at decreasing temperatures from 600 to 30 mK at 17 T. 
This technique enables us to optimize the concentration of the o-H$_{2}$ 
and to efficiently polarize the HD target within a shortened aging time. 
\end{abstract}

\pacs{Valid PACS appear here}
\maketitle

\section{Introduction}

We have been carrying out photoproduction experiments at the Laser Electron Photon 
beamline at SPring-8/LEPS since 2000~\cite{Nakano}. 
Linearly or circularly polarized photon beams in the energy range of 
1.5--3.0 GeV are produced by the backward Compton scattering of 
an ultraviolet laser from 8 GeV electrons~\cite{Mura2}. 
Photoproduction of various mesons and baryons, such as $\phi$~\cite{Mibe,Ryu}, 
$\pi$~\cite{Kohri3}, $K^{*}$~\cite{Hwang}, and 
hyperons~\cite{Zegers,Kohri1,Hicks,Mura,Kohri2}, was studied with 
unpolarized liquid hydrogen or deuterium targets. 
If a polarized nucleon target is introduced for the LEPS experiments, 
a new type of experiments can be realized to measure double-spin asymmetries that 
provide precious knowledge to the understanding of the hadron structure, 
its production mechanism, and the existence of exotic particles. 

Honig suggested the use of the hydrogen--deuteride (HD) molecule as a frozen-spin 
polarized target after the important work on the relaxation mechanism~\cite{Honig1}. 
The HD target had been developed at Syracuse~\cite{Honig2}, at 
the Brookhaven National Laboratory (BNL)~\cite{Rigney,Wei}, 
and at ORSAY~\cite{Rouille,Bassan}, and was 
used for physics experiments at the BNL~\cite{Hoblit} and 
the Jefferson Lab (JLAB)~\cite{Ho} by Sandorfi $et$ $al$. 
We started the development of the HD target at Osaka University 
in 2005~\cite{Kohri4,Utsuro,Yanai}. 

Given that the purity of commercially available HD gas is approximately 96\%, 
the HD gas is purified up to 99.99\% by a distiller~\cite{Ohta3}, and is analyzed 
by a gas chromatograph with a quadrupole mass spectrometer~\cite{Ohta2}. 
Pure H$_{2}$ gas with an amount of approximately 0.1\% is used as the catalyst, 
and is added to the purified HD gas to achieve long relaxation times. 
The HD gas is liquefied and solidified by liquid helium and the solid HD 
is cooled down to approximately 20-30 mK with a $^{3}$He-$^{4}$He dilution 
refrigerator. 
A 17 T magnetic field is generated by a superconducting solenoid, and is applied to 
build-up the HD polarization. 
The use of the static nuclear polarization at low temperatures and at a high-magnetic 
field strength for a three-month period leads to the generation and freezing of 
the HD spin polarization. 
The H$_{2}$ gas is composed of ortho-H$_{2}$ (o-H$_{2}$) with a spin 
of $J$ = 1 and para-H$_{2}$ (p-H$_{2}$) with a spin of $J$ = 0. 
In addition, the population ratio of o-H$_{2}$ to p-H$_{2}$ is 3:1 at room temperature. 
The o-H$_{2}$ molecules have a decay time of approximately 1 week at low 
temperatures, and generate approximately 2 $\mu$W heat at the beginning of the aging process. 
At the end of the aging period, almost all the o-H$_{2}$ molecules are converted 
to the p-H$_{2}$ molecules and engage in weak spin interactions with the HD. 
We obtained a relaxation time of approximately 8$\pm$2 months for the $^{1}$H nucleus, 
which was adequately long for the conduct of the planned experiments 
at SPring-8~\cite{Kohri5}. 
Once the spin polarization is frozen at Osaka University, the temperature can be increased 
and the magnetic field can be decreased. 
The HD target is transported to SPring-8 at a temperature of 1.5 K at a magnetic field of 
1 T for photoproduction experiments. 

In the past, the calibration of the $^{1}$H polarization was carried out based on 
Nuclear Magnetic Resonance (NMR) measurements with magnetic field sweeps at 
approximately 1 T with a frequency of 40 MHz at 4.2 K~\cite{Ohta1}. 
After the aging of the HD target at 17 T for 3 months, the magnetic field was 
decreased to 1 T and the H polarization was obtained based on the estimation of 
the ratio of the area of the final NMR signal to that of 
the calibration signal. 
A three-month aging period is very long and the consumption of liquid helium at 
a rate of approximately 24 L/day for the operation of 
the dilution refrigerator and the superconducting solenoid is costly. 
Most of the liquid helium can be supplied by the Low Temperature 
Center of Osaka University. 
To ensure smooth operations, we purchased commercial liquid helium for use 
during long holiday periods. 
Evaporated helium gas was returned to the Low Temperature Center for recycling. 

If the concentration of the o-H$_{2}$ is decreased at the beginning of the aging, 
the aging time can be shortened. 
The relaxation time of the solid HD depends on the concentration of the o-H$_{2}$, 
temperature, magnetic field and so on. 
The relaxation time of the solid HD was measured in the temperature range of 1.2--4.2 K 
at various concentrations of the o-H$_{2}$~\cite{Honig2,Bouchigny1,Bouchigny2,Bouchigny3,Hardy}. 
However, no measurements were conducted at temperatures lower than 1 K and 
at magnetic fields higher than 10 T. 
Although it was necessary to measure the relaxation time of the HD polarization 
within the temperature range below 1 K at 17 T for the optimization of the concentration of 
the o-H$_{2}$, there were technical difficulties in monitoring the build-up of the polarization 
by NMR measurements at high-frequencies of approximately 700 MHz. 
To overcome these difficulties, we developed a new NMR system which could be operated within 
a broad frequency range up to 726 MHz. 
Some brief explanations have been reported elsewhere~\cite{Ohta4}. 

\section{NMR systems}

\subsection{Portable NMR system with PCI eXtensions for instrumentation}

We measure the polarization of the $^{1}$H and $^{2}$H nuclei at Osaka 
University and at SPring-8. 
In order to make reliable calibration for the polarization, 
it was necessary to use the same NMR system at both sites. 
However, the weight of the conventional NMR system was 80 kg, and frequent transportation 
of the system (which was mounted on a rack with a height of 2 m) between Osaka University 
and SPring-8 was not easy. 
We constructed a portable NMR system with an operating software system with 
PCI eXtensions for Instrumentation (PXI)~\cite{Ohta1}. 
The weight of the portable NMR system was only 7 kg and the cost was 
reduced to 25\%. 

The portable NMR system consisted of PXI-1036 (chassis), 
PXI-8360 (connection between laptop PC and PXI), PXI-5404 (signal generator), and 
PXI-5142 (ADC), which were developed by the National Instruments Company. 
This system was controlled by a LabVIEW program on the laptop. 
The frequency range of the PXI-5404 ranged from 0 to 100 MHz, and was 
suitable for NMR measurements of the $^{1}$H nucleus with 
magnetic field sweeps at a field strength of approximately 1 T. 
The signal-to-noise (S/N) ratio of the portable NMR system depended on the performance 
of the laptop which was used to operate it. 

\subsection{New NMR system using frequency sweeps at high frequencies}

We developed a new NMR system that operated in a wide frequency range 
up to 726 MHz. 
Given that the polarization measurement was performed during the aging 
of the HD target at 17 T, the superconducting solenoid was operated with 
a persistent current mode, and a frequency sweep method was applied for the 
polarization measurements. 

\begin{figure}[h]
\begin{center}
\includegraphics[width=7.5cm]{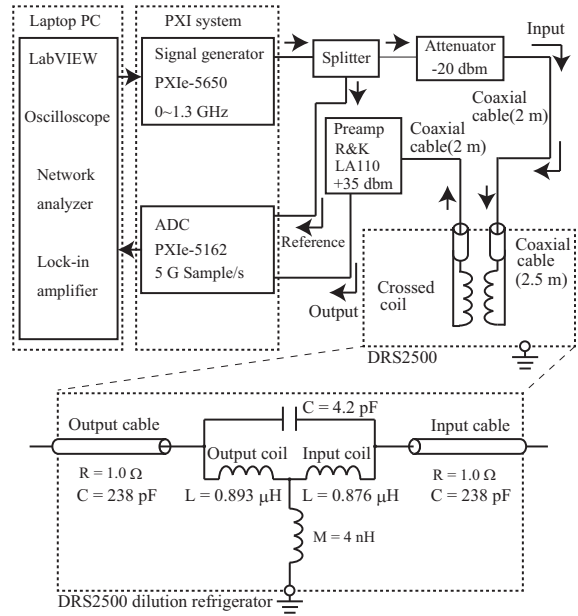}%
\caption{Schematic of the new NMR system. 
The crossed-coil method and frequency sweeps were applied. }%
\label{fig:newnmr}
\end{center}
\end{figure}

The signal generator PXI-5404 was replaced with PXIe-5650 which 
generated radiofrequency (RF) signals at frequencies up to 1.3 GHz. 
The analog-to-digital converter (ADC) PXI-5142 was also replaced with PXIe-5162 
which sampled the data at different rates up to 5 G Sample/s. 
Instead of the single-coil method used in the previous NMR system~\cite{Ohta1}, 
the crossed-coil method was applied, as shown in Fig.~\ref{fig:newnmr}. 
Given that the tuning circuit in the previous NMR system specified the frequency and 
required manual operations for each nucleus, the tuning circuit was not used. 
Although the measurements without the tuning circuit resulted in poor 
S/N ratios, automatic NMR measurements within wide frequency ranges 
were performed. 

\section{Experiments}

\subsection{Dilution refrigerator, target cell and NMR coils}

\begin{figure*}[htpb]
\includegraphics[width=14.0cm]{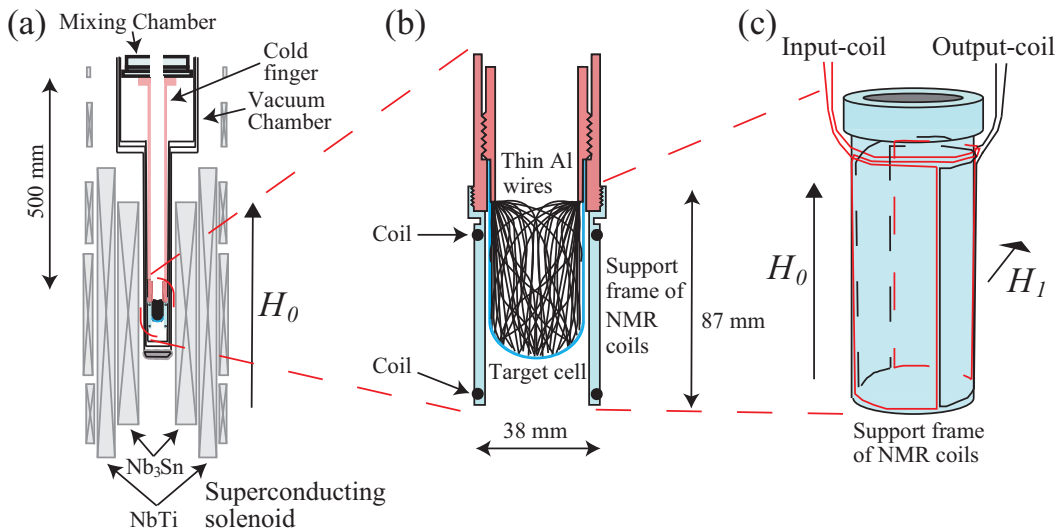}%
\caption{(a) Mixing chamber of the DRS2500 dilution refrigerator 
with the cold finger surrounded by the superconducting solenoid made 
of NbTi and Nb$_{3}$Sn. 
(b) Cross-section of the target cell and support frame 
of the NMR coils. 
(c) Structure of the support frame of the NMR coils with the directions 
of the magnetic field $H_{0}$ of the superconducting solenoid and the applied 
radiofrequency (RF) field $H_{1}$. }
\label{fig:cellcoil}
\end{figure*}

We used a $^{3}$He-$^{4}$He dilution refrigerator (DRS2500) produced 
by Leiden Cryogenics B.V.~\cite{Leiden} in the Netherlands to 
cool the HD target. 
The DRS2500 refrigerator has a lowest temperature of 6 mK and a cooling power of 
2500 $\mu$W at 120 mK. 
A strong magnetic field was produced by the superconducting solenoid 
(NbTi and Nb$_{3}$Sn) produced by JASTEC Co., Ltd.~\cite{Jastec} 
in Japan. 
The target cell was attached to a cold finger made of pure copper 
(99.99\%) with a length of 500 mm. In turn, the cold finger was attached 
to the mixing chamber with a lowest temperature, as shown 
in Fig.~\ref{fig:cellcoil}(a). 
A carbon resistance thermo sensor was used to measure the temperature 
of the mixing chamber. 

The HD target cell and the support frame of the NMR coils, shown 
in Fig.~\ref{fig:cellcoil}(b, c), were made of 
Kel-F (Poly-Chloro-Tri-Fluoro-Ethylene (PCTFE)) which did not contain 
any hydrogen. 
A Teflon-coated silver wire with a diameter of 0.3 mm was used for 
input signals. 
It was wound to form a single-turn saddle coil on the support frame. 
The other wire was also wound on the support frame in the perpendicular 
direction and served as the crossed-coil for picking up output signals. 
The RF field of $H_{1}$ was produced by the coil for input signals, and 
the direction of $H_{1}$ was perpendicular to that of the magnetic field 
$H_{0}$ produced in the superconducting solenoid. 
Thin aluminum wires (20\% in weight of the HD target) with a purity 
higher than 99.999\% were soldered on the target cell to insure 
the cooling of the solid HD. 

\subsection{Cooling HD target by dilution refrigerator}

The DRS2500 refrigerator and superconducting solenoid were precooled to 77 K 
by liquid nitrogen. 
After the liquid nitrogen was blown out, liquid helium cooled them to 4.2 K. 
The HD gas (1 mol), which had an o-H$_{2}$ impurity of 0.3\%, was liquefied at 
approximately 20 K and solidified at 4.2 K in the target cell. 
The superconducting solenoid was excited with a current of 270 A to produce 
a magnetic field strength of 17 T, and the operation was changed to 
the persistent current mode. 
NMR measurements with frequency sweeps for $^{1}$H, $^{2}$H, and $^{19}$F 
nuclei were initiated. 
The $^{1}$H and $^{2}$H nuclei were the main components of the HD target, 
and the $^{19}$F nucleus was contained in the target cell and support frame 
of the NMR coils. 
The NMR frequencies for the $^{1}$H, $^{2}$H, and $^{19}$F nuclei were 726, 111, 
and 683 MHz, respectively, at 17 T. 
The speed of the frequency sweeps was 0.544 MHz/s. 
We accumulated 100 k data points and estimated the average values at 
each frequency point. 

The temperature of the target decreased to 600 mK when the 1K pot of the DRS2500 was 
pumped and $^{3}$He gas was liquefied into the mixing chamber. 
When $^{4}$He gas was also liquefied into the mixing chamber, the temperature of 
the HD target became lower than 100 mK and gradually dropped down to 30 mK. 
A heater (power of 0.09 W) was applied to increase 
the flow of the circulating $^{3}$He gas. 

\subsection{Polarization}

The nuclear spins of $^{1}$H, $^{2}$H, and $^{19}$F nuclei are 1/2, 1, and 1/2, 
respectively. 
If the population distribution of the spin system obeys the Boltzmann statistics, 
the polarization of the $^{1}$H or $^{19}$F nucleus with the spin 1/2 can be 
calculated according to, 
\begin{equation}
\begin{split}
P_{~^{1}H/^{19}F} &= \frac{N_{+} - N_{-}}{N_{+} + N_{-}}\\
  &= {\rm tanh}(\frac{\mu H_{0}}{k_{B}T_{TE}}), 
\end{split}
\end{equation}
where $N_{+}$ and $N_{-}$ are the numbers of substates $m$=+1/2 and -1/2, respectively. 
$H_{0}$ is the magnetic field, $\mu$ is the magnetic moment of 
the $^{1}$H (2.793$\mu_{N}$) or $^{19}$F (2.629$\mu_{N}$) nucleus, 
$\mu_{N}$=3.152$\times$10$^{-8}$ eV/T, 
$k_{B}$ is the Boltzmann constant (8.617$\times$10$^{-5}$ eV/K), 
and $T_{TE}$ is the temperature of the thermal equilibrium state. 

In the case of the $^{2}$H nucleus with the spin 1, the vector polarization 
is calculated to be 
\begin{equation}
\begin{split}
P_{~^{2}H} &= \frac{N_{+} - N_{-}}{N_{+} + N_{0} + N_{-}} \\
         &= \frac{4{\rm tanh}(\frac{\mu_{D} H_{0}}{2k_{B}T_{TE}})}{3 + {\rm tanh}^{2}(\frac{\mu_{D} H_{0}}{2k_{B}T_{TE}})},
\end{split}
\end{equation}
where $N_{+}$, $N_{0}$, and $N_{-}$ are the numbers of substates 
$m$=+1, 0, and -1, respectively, and $\mu_{D}$ is the magnetic moment 
of the $^{2}$H nucleus (0.857$\mu_{N}$). 

The polarizations of the $^{1}$H, $^{2}$H, and $^{19}$F nuclei at the 
thermal equilibrium state at 17 T are calculated to be 0.41\%, 0.08\%, and 0.39\% 
at 4.2 K, and 52\%, 12\%, and 50\% at 30 mK, respectively. 
Given that the magnetic moment of $^{19}$F is close to and slightly 
smaller than that of $^{1}$H, these two nuclei have 
similar polarizations. 
The $^{2}$H nucleus has a smaller magnetic moment than those of 
the $^{1}$H and $^{19}$F nuclei, and yields a smaller $^{2}$H polarization.

\section{Results}

\subsection{NMR signals}

\begin{figure}[h]
\begin{center}
\includegraphics[width=8.5cm]{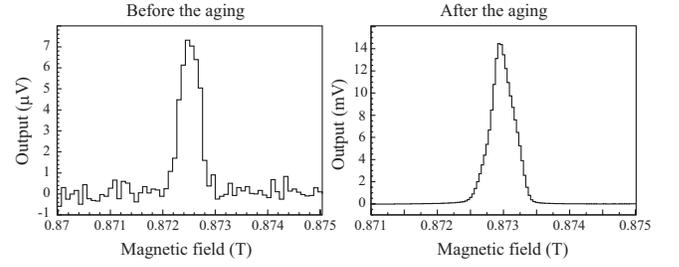}%
\caption{ NMR signals of the $^{1}$H nucleus measured with the 
portable NMR system before and after the aging process.}
\label{fig:oldsignals}
\end{center}
\end{figure}

Figure \ref{fig:oldsignals} shows NMR signals of the $^{1}$H nucleus 
measured before and after the aging process based on the use of the portable 
NMR system. 
The NMR signals before the aging were measured at 4.2 K and 1 T. 
The NMR signals of the $^{1}$H nucleus were measured again at 0.3 K and 1 T 
for comparison after the aging process at temperatures of about 20 mK 
at 17 T over a three-month period. 
The NMR signals measured after the aging were approximately 2000 times larger than 
those measured before the aging process. 

\begin{figure}[h]
\begin{center}
\includegraphics[width=8.5cm]{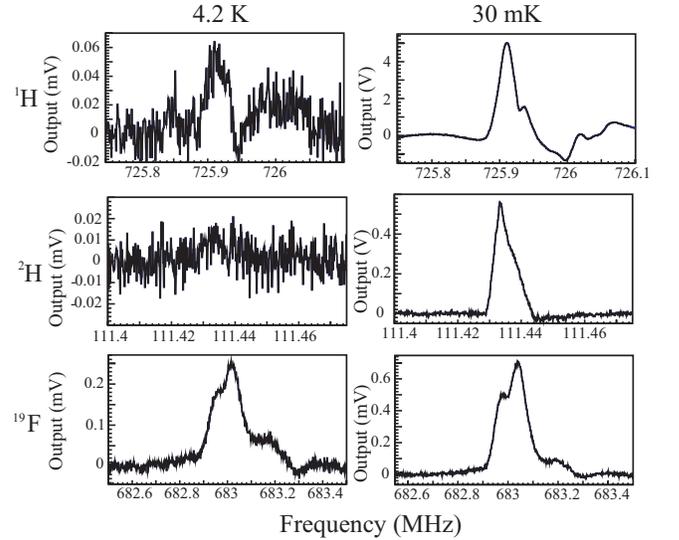}%
\caption{NMR signals of the $^{1}$H, $^{2}$H, and $^{19}$F nuclei 
measured at 4.2 K (left) and 30 mK (right) at 17 T with the 
new NMR system (Fig.~\ref{fig:newnmr}). }
\label{fig:signals}
\end{center}
\end{figure}

NMR signals measured with frequency sweeps at 4.2 K and 30 mK at 17 T 
based on the use of the new NMR system (Fig.~\ref{fig:newnmr}) are 
shown in Fig.~\ref{fig:signals}. 
The signals of $^{1}$H and $^{2}$H nuclei were small and the S/N ratios 
were poor at 4.2 K. 
Given that the number of $^{19}$F nuclei was much larger than those of 
the $^{1}$H and $^{2}$H nuclei, and given that the $^{19}$F nuclei were 
located near the NMR coils where the coil sensitivity was high, 
the $^{19}$F signals were clearly observed even at 4.2 K. 
The intensities of the $^{1}$H and $^{2}$H signals became larger by 
approximately 100 and 30 times, respectively, when the temperature 
decreased from 4.2 K to 30 mK. 
The intensities of the $^{19}$F signals increased by a factor of approximately 
3 at 30 mK. 
It was considered that the deterioration of the $^{1}$H and $^{19}$F signals 
at 30 mK was caused by the high-frequency signal detection difficulties. 
The $^{1}$H and $^{19}$F signals were clearly observed at magnetic fields 
below 7 T, however these deteriorated at field strengths above 7 T. 

\subsection{Monitoring the build-up of the polarization}

\begin{figure}[h]
\begin{center}
\includegraphics[width=7cm]{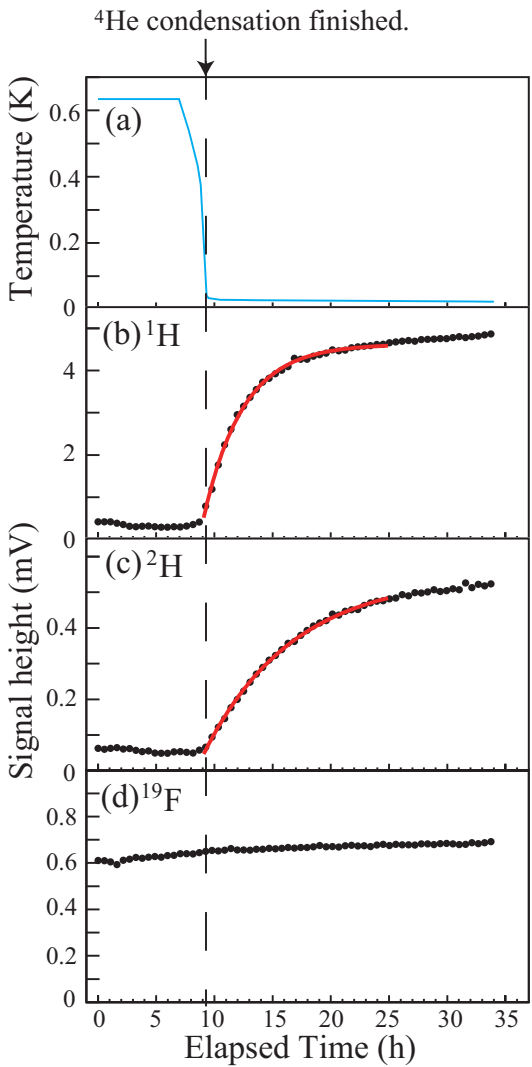}%
\caption{(a) Temperature of the mixing chamber measured by 
a carbon resistance thermometer. 
The NMR signal heights of (b) $^{1}$H, (c) $^{2}$H, and (d) $^{19}$F nuclei 
at the beginning of the aging of the 
HD target at 17 T. The solid curves are the results of the fits obtained 
with the use of function (3). }
\label{fig:monitor}
\end{center}
\end{figure}

Only the central regions of the NMR imaginary signals in Fig.~\ref{fig:signals} 
were fitted with a Gaussian function, and the signal heights were obtained. 
The temperature of the mixing chamber and the build-up of 
the polarization of the $^{1}$H, $^{2}$H, and $^{19}$F nuclei 
are shown in Fig.~\ref{fig:monitor}. 
We succeeded in monitoring the build-up of the polarization during 
the aging process of the HD target at 17 T. 

The $^{3}$He condensation was completed at 0 h, while the 
condensation $^{4}$He was initiated at 2.5 h and was completed at 9 h. 
The temperature of the mixing chamber decreased from 600 to 30 mK 
during the $^{4}$He condensation. 
The $^{1}$H polarization started to grow at 9 h. 
The polarization grew up to its maximum value within 1 day. 
The $^{2}$H polarization also started to grow at 9 h. 
The speed of the growth of the $^{2}$H polarization was slower 
than that of the $^{1}$H polarization. 
Both the $^{1}$H and $^{2}$H polarizations became approximately 10 times 
larger at 30 mK than those at 600 mK. 
The $^{19}$F polarization became larger by only 10\% when 
the temperature decreased from 600 mK to 30 mK. 
Given that the NMR signals of $^{19}$F at 30 mK are larger than 
those at 4.2 K by approximately 3 times, the actual temperature 
was estimated to be approximately equal to 1.5 K. 
The thermal conductivity of Kel-F at 500 mK was approximately 
2$\times$10$^{-5}$ W/cm$\cdot$K, and was adequately large for cooling 
the NMR support frame~\cite{Anderson}. 
Insufficient cooling of the NMR support frame was inferred owing to 
the poor thermal conductivity between the support frame and the cold finger. 
Grease should be added to the screws for better thermal conductivity 
in the next cooling attempt. 

The NMR signal heights within the range of 9 to 25 h were fitted by the function, 
\begin{equation}
P = P_{0}(1-\exp(- \frac{t-t_{0}}{T_{1}})), 
\end{equation}
where $P_{0}$, $t_{0}$, and $T_{1}$ are free parameters. 
The $T_{1}$ values of the $^{1}$H and $^{2}$H nuclei at 30--600 mK 
at 17 T were 2.96$\pm$0.03 and 7.72$\pm$0.72 h, respectively. 
By fitting the data with narrow regions, the $T_{1}$ value of the $^{1}$H nucleus 
increased as time elapsed. 
Given that the concentration of o-H$_{2}$ did not decrease so fast, 
it was considered that the prolonged value of $T_{1}$ was inferred to be 
caused by the low temperature. 

We also carried out data analyses based on the areas of the NMR peak regions 
which were estimated by integration. 
As a result of this analysis, the relaxation times of the 
$^{1}$H and $^{2}$H nuclei were 3.69$\pm$0.03 and 7.90$\pm$0.12 
h, respectively. 
The uncertainties due to the selection of the integration range were about 
0.22 and 0.03 h for the $^{1}$H and $^{2}$H nuclei, respectively. 

The $^{2}$H relaxation time was longer compared with the $^{1}$H relaxation time. 
In previous studies, the same results were obtained 
at 1.8 K and 0.85 T~\cite{Bouchigny3}. 
At the beginning of the aging process, we used H$_{2}$ as a catalyst but did not 
use D$_{2}$ in the HD gas, which may have led to a longer $^{2}$H relaxation time. 

The NMR data were measured 12 days after the liquefication and 
solidification of the HD. 
The o-H$_{2}$ concentration was estimated to decrease from 0.3\% 
to 0.05\% during the measurements. 
Given that the relaxation times of the $^{1}$H and $^{2}$H nuclei were 
found to be short enough, the o-H$_{2}$ concentration of 0.05\% should be 
reduced to shorten the aging time of the HD target. 

\section{Summary and future outlook}

In order to optimize the amount of o-H$_{2}$ in the HD target and the aging time, 
we developed a new NMR system, and succeeded in the monitoring of the build-up 
of the polarizations of the $^{1}$H and $^{2}$H nuclei at 17 T. 
The polarizations were found to grow within 1 day when the temperature 
decreased from 600 to 30 mK. 
The relaxation times of the $^{1}$H and $^{2}$H nuclei at 30--600 mK and 
17 T were obtained as 2.96$\pm$0.03(stat)$\pm$0.73(syst) and 
7.72$\pm$0.72(stat)$\pm$0.18(syst) h, respectively, where 
the differences between the results of two different analyses are 
considered as the systematic uncertainties. 
The o-H$_{2}$ concentration of 0.05\% was excessively large for the build-up 
of the polarizations. 
Accordingly, in future work, we will optimize the concentration of o-H$_{2}$, 
and shorten the traditionally used three-month aging period. 
In addition, the present frequency sweep method will be useful 
for the monitoring of the polarization of the HD target during the 
photoproduction experiments at SPring-8.

In these experiments, we also observed the NMR signals of the $^{27}$Al and 
$^{35}$Cl nuclei. 
Recently, we started developing a polarized $^{139}$La target for 
a T-violation experiment with polarized neutron beams at J-PARC. 
The present NMR system can be used for a broad range of frequencies, 
including the frequency of 102 MHz for the $^{139}$La nucleus at 17 T. 
Although the maximum frequency in this experiment was 726 MHz, 
the present NMR system can generate and observe signals at higher frequencies 
up to 1.3 GHz. 
The present technique would play important roles not only 
for the target development of HD but also for various other polarized 
nuclear target developments. 

\begin{acknowledgments}
The presented study involved the conduct of experiments at the BL33LEP of 
SPring-8 with the approval of the Japanese Synchrotron Radiation 
Research Institute (JASRI) as the contract beam line 
(Proposal No. BL33LEP/6001). 
We are grateful to the staff of the Low Temperature Center of 
Osaka University for supplying us with the required liquid helium. 
We thank Dr. J.-P. Dideletz, Dr. S. Bouchigny, Dr. G. Rouille, 
Professor G. Frossati, Dr. N. R. Hoovinakatte, Dr. A. M. Sandorfi, 
Dr. X. Wei, Dr. M. M. Lowry, and Dr. T. Kageya 
for their important advice. 
We also thank Professor K. Fukuda, Dr. T. Kunimatsu, and Professor M. Tanaka, 
for the construction of the primary NMR system and for the provision of 
some additional modules. 
The present work was supported in part by the Ministry of Education, 
Science, Sports, and Culture of Japan, and 
by the National Science Council of the Republic of China. 
\end{acknowledgments}

\end{document}